\documentclass[conference]{IEEEtran}

\usepackage{cite, graphicx}
\usepackage{hyperref}
\usepackage[cmex10]{amsmath}
\usepackage{amssymb}
\usepackage{siunitx}
\ifCLASSOPTIONcompsoc
\usepackage[caption=false,font=normalsize,labelfont=sf,textfont=sf]{subfig}
\else
\usepackage[caption=false,font=footnotesize]{subfig}
\fi
\usepackage{algorithmic}
\usepackage{multirow}

\begin{document}
\title{HyCell: Enabling GREEN
Base Station Operations in Software-Defined
Radio Access Networks}
\author{\IEEEauthorblockN{Tao Zhao,
Liumeng Wang, Xi Zheng, Sheng Zhou, and Zhisheng Niu}
\IEEEauthorblockA{%
Tsinghua National Laboratory for Information Science and Technology (TNList)\\
Department of Electronic Engineering, Tsinghua University, Beijing 100084, China\\
\{t-zhao12, wlm14, zhengx14\}@mails.tsinghua.edu.cn, \{sheng.zhou, niuzhs\}%
@tsinghua.edu.cn}}
\maketitle

\begin{abstract}

  The radio access networks (RANs) need to support
  massive and diverse data traffic with limited spectrum and energy.
  To cope with this challenge, software-defined radio access network (SDRAN)
  architectures have been proposed to renovate the RANs. However, current
  researches lack the design and evaluation of network protocols. In this paper,
  we address this problem by presenting the protocol design and evaluation of
  hyper-cellular networks (HyCell), an SDRAN framework making base station (BS) operations globally
  resource-optimized and energy-efficient (GREEN).
  Specifically, we first propose a separation scheme to realize the
  decoupled air
  interface in HyCell. Then we design a BS dispatching protocol which determines and assigns
  the optimal BS for serving mobile users, and a BS sleeping protocol
  to improve the network energy efficiency. Finally, we evaluate
  the proposed design in our HyCell testbed. Our evaluation validates the feasibility of the
  proposed separation scheme, demonstrates the effectiveness of BS dispatching,
  and shows great potential in energy saving through BS sleeping control.
\end{abstract}

\begin{IEEEkeywords}
  Software-defined networking, radio access network, base station sleeping.
\end{IEEEkeywords}

\section{Introduction}
\label{sec:intro}

The radio access networks (RANs) are facing great
challenges in the mobile Internet era.
On one hand, next-generation RANs are expected to support 1000 times increased
data traffic~\cite{4gamericas2014meeting} with limited spectrum and energy.
Therefore, both spectrum efficiency and energy efficiency should be
vastly improved.
On the other hand, emerging applications and services put forward increasingly diverse
requirements for connections in terms of capacity, latency, and reliability.
Hence, next-generation RANs need to be sufficiently flexible to accommodate
evolving applications.

To tackle these challenges, software-defined networking (SDN) have
been proposed to renovate RANs.
SoftRAN~\cite{gudipati2013softran} introduced a logically centralized big base
station (BS) to globally optimize the RAN resources so that the
network efficiency can be improved.
The control-data separation concept of SDN was extended to decouple the control and
data coverages at the
air interface in RANs~\cite{niu2012energy,capone2012looking,ishii2012novel}.
With the decoupled air interface, BS sleeping control can be effectively employed to
reduce energy waste by adapting to real traffic dynamics and
improve the network energy efficiency without generating coverage holes.
Moreover, CONCERT~\cite{liu2014concert} proposed to deploy RANs as
software-defined services, which can dramatically improve the flexibility of
BS operations.
However, these studies focus on the architecture design only, and
leave the protocol design as an open research issue.

Besides design, it is of much research interest to evaluate the SDN concepts in
practical RAN
implementations.
Our previous work presented a prototype system
which demonstrated the feasibility of decoupled air interface on top of
the GSM standard~\cite{zhao2013software}.
But only a single type of service (namely voice calls) was investigated, and no
attempt to realize dynamic BS operations was made.
PRAN~\cite{wu2014pran} showed a preliminary implementation of dynamic resource
allocation of BSs with centralized RAN schedulers. However, other BS operations such as BS
sleeping were not studied.
To our knowledge, there are few efforts to evaluate the
protocols of BS operations in software-defined RANs (SDRANs) with practical
implementations.

In this paper, we present the protocol design and evaluation of
our SDRAN framework named hyper-cellular networks (HyCell). HyCell enables
globally resource-optimized and energy-efficient (GREEN~\cite{niu2011tango}) BS operations
by exploiting the decoupled air
interface, centralized BS control, and software-defined BS functions.
To design the decoupled air interface,
we take an evolutionary approach
and propose our separation scheme for current 3GPP standards, which
is beneficial for network migration.
Based on that, we design our BS dispatching protocol,
which determines and assigns the globally optimal BS to serve the user
requests, as well as our BS sleeping protocol.
Moreover, we prototype a HyCell testbed on a software-defined radio (SDR)
platform, and use it to evaluate our design.
The main contributions of our work are summarized as follows:
\begin{enumerate}
  \item We propose a separation scheme to realize the decoupled air interface
    for existing 3GPP
    standards from the aspects of
    both network functionalities and logical channels, and demonstrate its
    feasibility in the testbed.
  \item We design a BS dispatching protocol for global optimization of
    network resources, and
    implement a BS dispatching scheme to effectively achieve
    load balancing among multiple BSs.
  \item We design a BS sleeping protocol to
    improve the network energy efficiency,
    and present an implementation with a threshold-based algorithm, which
    shows about 60\% energy saving gain in our testbed.
\end{enumerate}

The rest of this paper is organized as follows.
We first give an overview of the HyCell architecture as well as
the challenges and solutions to realize it in Section~\ref{sec:overview}.
Then we present the design of the separation scheme, BS
dispatching, and BS sleeping in Section~\ref{sec:design}.
Evaluation of the testbed implementation is given in Section~\ref{sec:eval}.
Section~\ref{sec:con} concludes the paper.

\section{Overview}
\label{sec:overview}
\subsection{Architecture}
\label{sec:arch}
\begin{figure}[!t]
  \centering
  \includegraphics[width=0.48\textwidth]{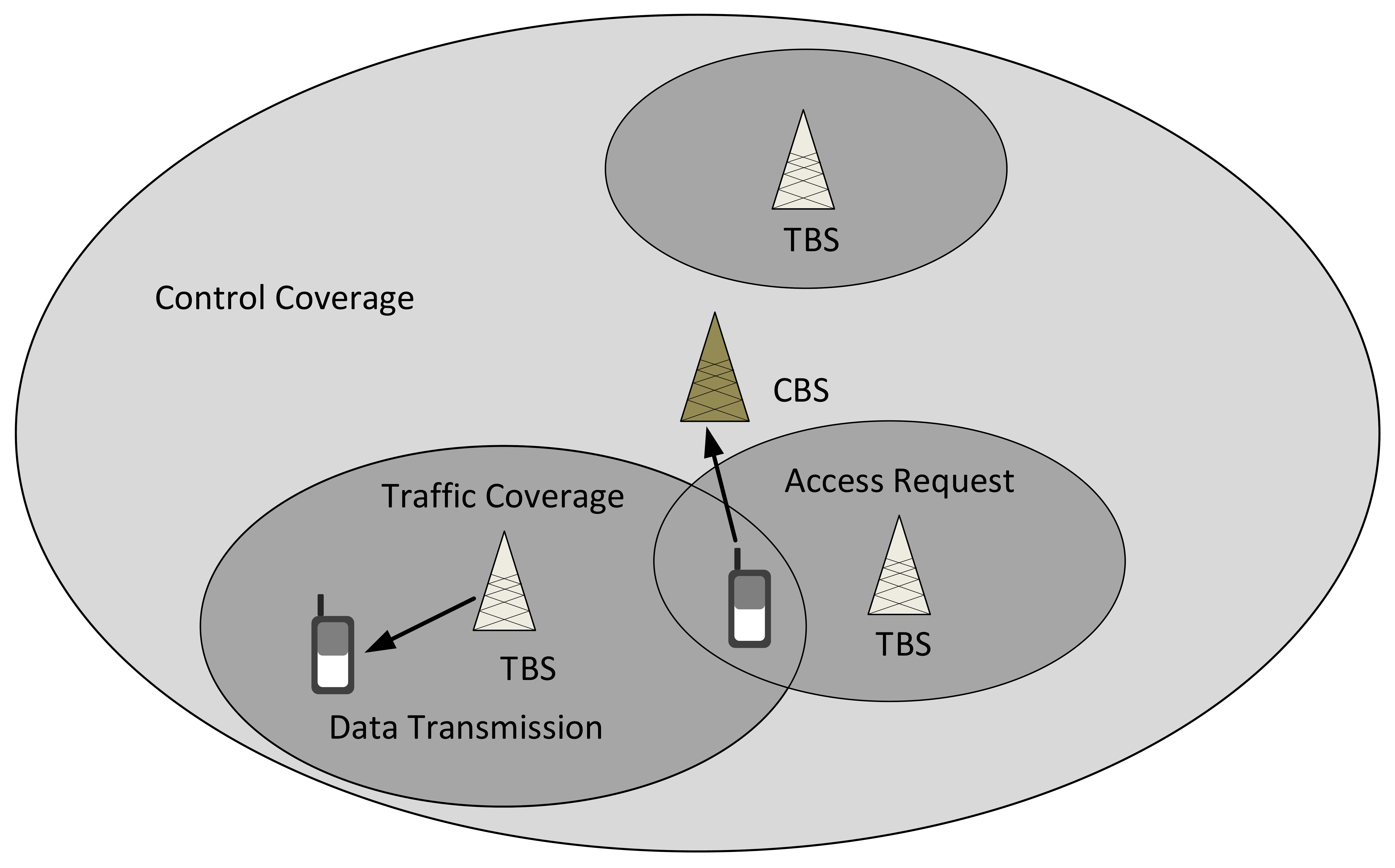}
  \caption{HyCell architecture.}
  \label{fig:arch}
\end{figure}

As illustrated in Fig.~\ref{fig:arch}, HyCell decouples the air interface of the
RAN by separating the control coverage from the traffic coverage with two types of BSs.
Control BSs (CBSs) provide the control coverage, while traffic BSs (TBSs)
provide the traffic coverage.
Typically CBSs have large coverage areas, within which
multiple traffic BSs (TBSs) are deployed.
CBSs provide the network control to underlying TBSs and mobile users.
In particular, they grant the network access to mobile users.
To guarantee a basic level of data service and to cope with high mobility users,
CBSs can also be used to provide low-rate data services such as voice call service to the UE.
Unlike CBSs, TBSs are only responsible for high-rate data services.
Specifically, there can be different subtypes of TBSs
to support different classes of high-rate data services.
Through the separation, CBSs and TBSs can be more simplified than
conventional BSs.

Under the separation architecture, CBSs provide centralized control,
and enable dynamic operations of TBSs. CBSs and TBSs serve mobile users
collaboratively.
When the user equipment (UE) is powered on,
it searches for nearby CBSs and registers to the network through a CBS.
CBSs gather network state information such as UE locations and traffic load
from the mobile users and TBSs underlying its coverage,
and thus hold a global view of the network.
When high-rate data services are required, the UE sends requests to the CBS
and the CBS dispatches one or more TBSs for the high-rate data transmission afterwards.
When the network traffic load is light, TBSs can be turned off under the
command of the CBS to reduce the
energy consumption of the network while the CBS preserves the network coverage.

Moreover, HyCell decouples the software which realizes the BS functions from the
hardware which converts signals between baseband and radio frequency.
In this way, BS functions are software defined by high-level programming languages, and
it is much easier to update the BSs to handle
new mobile Internet applications and services. Further, with the help of
virtualization, BSs can become virtual instances in cloud data centers.
Network efficiency can be improved with resource pooling, and BS operations can
be more flexible with the help of virtual machine migration.

With the decoupled air interface, centralized BS control, and
software-defined BS functions,
HyCell enables GREEN BS operations in SDRANs.
It leads to a promising path towards next-generation RANs.

\subsection{Challenges and Solutions}
\label{sec:cha}

We summarize the challenges of protocol design in HyCell and our proposed
solutions as follows.

\paragraph{Separation of Air Interface}
In this paper, we target at current air interface protocols and propose our separation
scheme to design a decoupled air interface.
However,
it is a daunting task to analyze the air interface in existing 3GPP standards.
The physical layer signals are difficult to categorize and the
interactions among them are complicated, making it difficult to separate the
control coverage from the traffic coverage.
To tackle this challenge, we propose our separate scheme from the aspects of network
functionalities and logical channels, rather than physical layer signals.
In our design, we jointly consider the two levels and ensure that the expected
network functionalities of CBSs and TBSs can be mapped to their
corresponding logical channels. This guides us to an effective separation
scheme of the air interface.

\paragraph{CBS-TBS Protocol Design}
The CBS-TBS protocols do not exist in current 3GPP standards.
However, the protocols are crucial for the SDRANs to optimize the
network performance. Specifically, we need a protocol to find the best TBSs and
assign them to serve the UE. We also need a protocol to dynamically switch off TBSs
to reduce energy waste and improve network energy efficiency.
The signaling interactions in the protocols must carry sufficient network
state information for global optimization. At the same time, it should
ensure realtime decision and accommodate potential
updates. To meet the requirements, we choose simple but effective metrics of
the network state information in message exchange. We also make our protocols
extensible by adopting modular design.

\paragraph{UE Transparency} When designing HyCell, we would like to guarantee
that the updates at the BS side are transparent to the UE side.
With UE transparency, the existing mobile terminals can work with
HyCell automatically. It brings the benefit of compatibility,
which is appealing during the protocol evolution and network upgrade. However, it also limits the
degree of freedom we have to realize the SDRAN. To achieve UE transparency, we carefully design
the separation scheme of the air interface and make sure that the UE-facing
interfaces are preserved. In addition, we exploit the backhaul connection
for CBS-TBS communication and cooperation, which optimizes the network without
the need of adding functions to mobile terminals.

\section{Key Design Aspects}
\label{sec:design}

In this section we describe our key design aspects.  First we analyze
the air interface of 3GPP standards and propose a separation scheme for HyCell.
Then a BS dispatching protocol design is proposed, and we also
present a BS sleeping protocol which achieves great energy saving gain without
sacrificing mobile users' quality of service.

\subsection{Separation of Air Interface}
\label{sec:sep}
\begin{table}[!t]
  \renewcommand{\arraystretch}{1.3}
  \caption{Logical channel separation of 3GPP standards.}
  \label{tab:sep}
  \centering
  \begin{tabular}{c|c|c|c|c}
    \hline
    \multicolumn{3}{c|}{Logical Channel} & \multicolumn{2}{c}{Separation} \\
    \hline
    GSM/GPRS & UMTS & LTE & CBS & TBS \\
    \hline
    BCH & BCCH & BCCH & \checkmark & \\
    \hline
    CCCH & CCCH, PCCH & CCCH, PCCH & \checkmark & \\
    \hline
    PACCH & DCCH & DCCH & & \checkmark \\
    \hline
    TCH & CTCH & MTCH, MCCH & \checkmark & \\
    \hline
    PDTCH & DTCH & DTCH & & \checkmark \\
    \hline
  \end{tabular}
\end{table}

First, we present our separation scheme of the air interface
from the view of network functionalities.
We categorize network functionalities into five classes: synchronization, broadcast of system
information, paging, low-rate data transmission, and high-rate data
transmission. To realize the decoupled air interface of HyCell, we need to
separate the network functionalities into two types of BSs. In our design,
CBSs are in charge of synchronization,
broadcast of system information, paging, and low-rate data transmission,
while TBSs are only responsible for high-rate data transmission.

Next, we propose our separation scheme from the perspective of logical channels
considering their corresponding network functionalities.
The logical channels of 3GPP standards can be classified into five groups,
while the specific channel names vary in different standards~\cite{sauter2011from}.
As shown in Table~\ref{tab:sep},
BCCH (broadcast control channel, BCH in GSM/GPRS) and
CCCH (common control channel, including PCCH for paging in UMTS and LTE)
are reserved by CBSs since they are in charge of network functionalities
including synchronization, paging, and broadcast of system information.
CTCH (common traffic channel, TCH in GSM/GPRS, MTCH and MCCH for multicast in LTE)
is also left at the CBS side to take care of low-rate data transmission.
At the TBS side are DTCH (dedicated traffic channel) and DCCH (dedicated
control channel), which are responsible for high-rate data transmission.
In the case of GSM/GPRS, we use PDTCH (packet data traffic channel) and
PACCH (packet associated control channel) for packet data transmission
and associated control.

Compared with the separation scheme of the GSM standard~\cite{zhao2013software},
here we differentiate the low-rate data service from the high-rate data service,
which results in a distinct separation scheme of logical channels.
This more fine-grained separation
allows the network to ensure a basic level of data service to the mobile users,
and thus is more practical.
Unlike the functionality separation scheme in existing
work~\cite{xu2013functionality}, we also present the separation scheme from the
aspect of logical channels. Besides, because of the joint
consideration of network functionalities and logical channels,
synchronization in the air interface resides only at the CBS side in our design.
We will show in Section~\ref{sec:eval}
that our separation scheme is feasible as long as fine synchronization
between CBSs and TBSs is guaranteed.

\subsection{Base Station Dispatching}
\label{sec:disp}

Through the separation of the air interface, the control coverage and the traffic
coverage of the RAN are decoupled. The CBS can gather the information of all
mobile users and TBSs in its coverage, and thus have a global view of the
network. Based on that, the CBS can dispatch one or more TBSs to serve the
high-rate data service request of mobile users in
a globally optimal fashion. To this end, a protocol of BS
dispatching is needed.

One possible design of BS dispatching is based on channel state information
(CSI) between the UE and the TBSs. If we collect the CSI and make it available
to the CBS, it can calculate the achieved data rate with each TBS, and choose
the TBS with the highest value. However, it is difficult to acquire the CSI
since the TBSs might have no prior connection to the UE. One possible solution
is that we embed pilot symbols in the channel requests. Then we
demand that TBSs monitor the channel requests and report the measured CSI to the
CBS. But besides the potential breakage of UE transparency, large quantity of CSI
also poses a heavy
burden on the backhaul link, for instance,
the X2 interface in LTE, especially in multiple antenna scenarios.

A more practical design is based on traffic load.
The load information can be readily acquired at each TBS, and aggregated at the
CBS. It is also an important metric to characterize the network state.
In a heavily-%
loaded network, balancing the load of multiple TBSs is beneficial because it
can potentially deliver better service in addition to improving radio resource
utilization and reducing the risk of system failure.\footnote{In the lightly-%
loaded scenario, BS sleeping is preferable, and the protocol will be described in
Section~\ref{sec:sleep}.}

\begin{figure}[!t]
  \centering
  \includegraphics[width=0.48\textwidth]{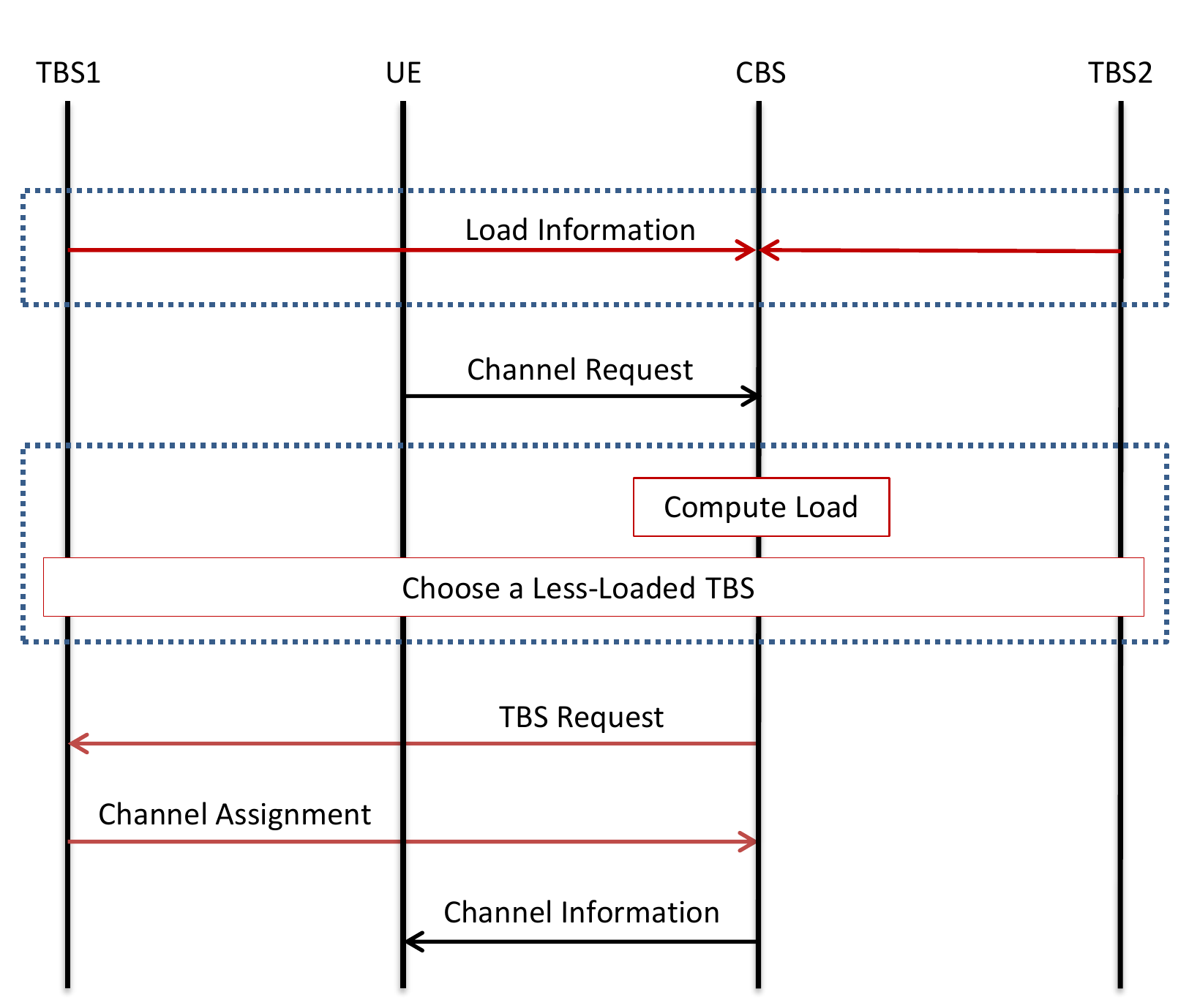}
  \caption{Signaling interaction of BS dispatching.}
  \label{fig:disp}
\end{figure}

Fig.~\ref{fig:disp} illustrates the signaling
interaction of the load-balancing BS dispatching protocol.
When the TBSs are actively serving the mobile users, they send to the CBS a
TBS load information message in packets through the backhaul link.
The packet header contains a label to indicate the message type, the TBS ID,
and the timestamp of packet transmission.
The packet payload is related to the specific air interface. For example, with
the decoupled GSM/GPRS air interface, the payload includes
the time slot usage, the absolute radio-%
frequency channel number (ARFCN) usage, and the data transmission rate.
After receiving the message, the CBS reserves the
statistical data for a preset period. When a new channel request of high-rate data
service arrives, the CBS computes the load of each TBS, and chooses the
least loaded TBS to provide the service. Then the CBS sends a TBS request
message to the chosen TBS, and waits for a response message with new channel
assignment. After that the CBS informs the user of the assigned channel, and
the link between the TBS and the mobile user is established.

\subsection{Base Station Sleeping}
\label{sec:sleep}

With the decoupled air interface, CBSs can provide an ``always-on''
control coverage, and we can turn TBSs into sleep mode flexibly and dynamically
without generating unwanted coverage holes. Because the CBS hold the global
network state information in its coverage, it can decide
which TBSs go to sleep to improve the network energy efficiency.

Next we introduce the protocol of BS sleeping in HyCell.
Assume the network load is light, and not
all TBSs are needed to meet the mobile users' service requirement.
The CBS periodically executes a BS sleeping control routine which determines
the modes of the underlying TBSs to maximize the network energy efficiency.
In each cycle,
it sends a BS sleeping message to each active TBS to be turned off,
and a BS waking up message to each sleeping TBSs to be turned on.
The BS sleeping and waking up messages are transmitted through the
backhaul link between CBSs and TBSs. They require acknowledgment to ensure
robust control of BS sleeping.

Thanks to the software-defined nature of BS functions,
We can adopt modular design, and make the API of the BS sleeping control routine
independent of the specific algorithm inside. In this way,
the algorithm involved can be updated without changing the overall protocol,
and thus the protocol is extensible.
Several BS sleeping algorithms have been developed in the
literature~\cite{zhou2009green,guo2013optimal,soh2013energy,zhang2014energy}.
Existing centralized algorithms for conventional RANs
can be put into use in the SDRANs by assuming the controller
to be at the CBS.
While centralized solutions is favored because of the optimality guarantee and
the ease to develop and implement, it might not suit all
needs. For instance, when TBSs are much densely deployed compared with CBSs,
the centralized decision could result in prohibitively high complexity. In such
case, decentralized control or distributed algorithms can be utilized to
reach a better tradeoff.

\section{Evaluation}
\label{sec:eval}
\subsection{Testbed Setup}
\label{sec:testbed}

\begin{figure}[!t]
  \centering
  \includegraphics[width=0.4\textwidth]{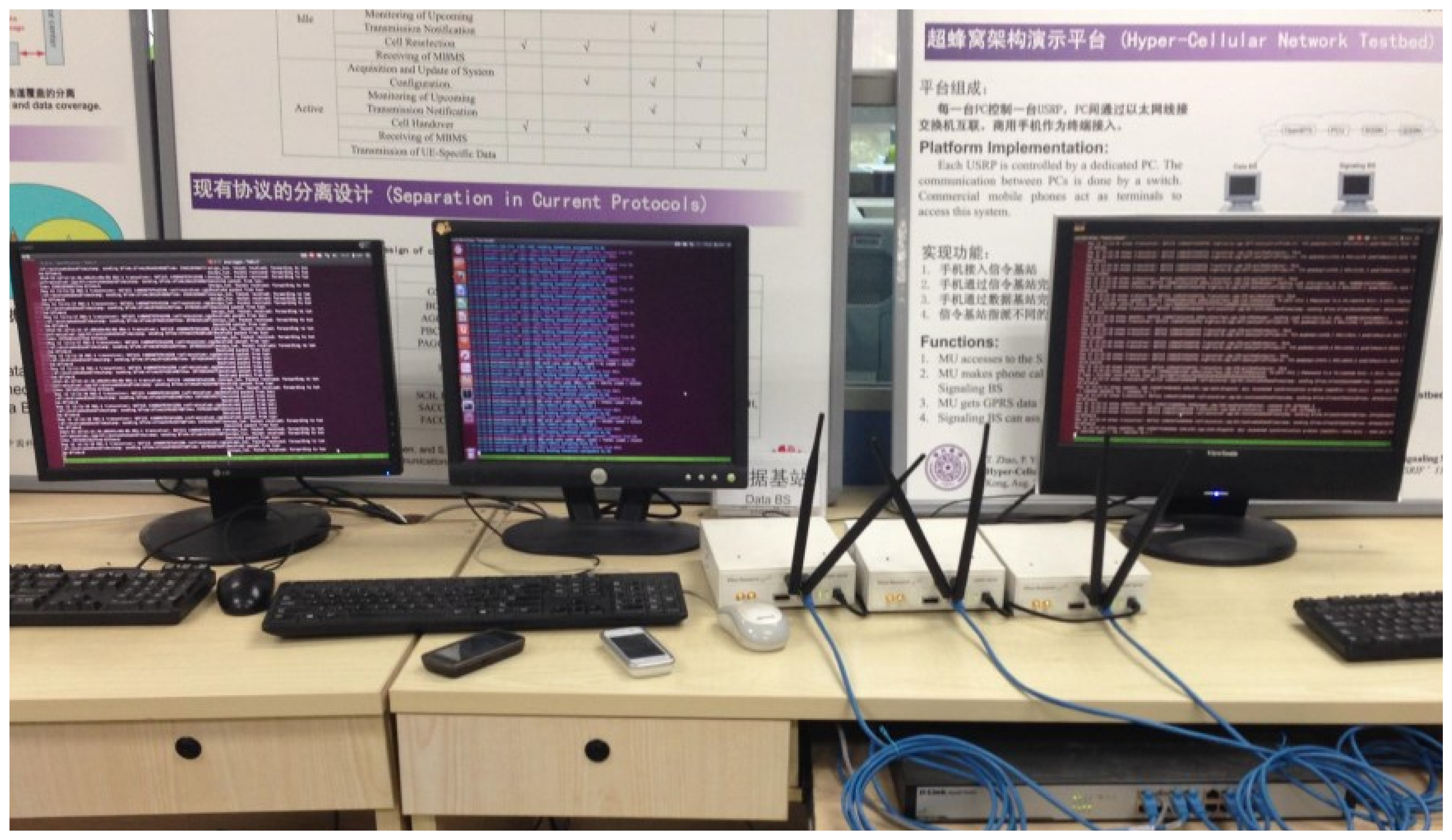}
  \caption{HyCell testbed.}
  \label{fig:testbed}
\end{figure}

As shown in Fig.~\ref{fig:testbed}, we utilize an SDR platform to prototype the
HyCell testbed, which can demonstrate GREEN BS operations in SDRANs.
In this testbed, we
employ three commodity PCs to run the software applications of one CBS and two TBSs.
The relevant parameters are listed in Table~\ref{tab:pc}.
We use two commercial mobile phones (Nokia 5230) to test the network.
In our tests, the two TBSs have overlapping traffic coverage, where a mobile user can
be served by either TBS as in Fig.~\ref{fig:arch}.
Note that it is straightforward to add more TBSs to the testbed by setting up
more PCs with the TBS software.

We use USRP N210s with WBX daughter boards and VERT900 antennas
from Ettus Research\footnote{\url{http://www.ettus.com}}
as the hardware to transform data
between radio frequency and baseband.
The USRP N210s and PCs are all connected to a Gigabit Ethernet switch.
We install the GPS-disciplined oscillators (GPSDO) module to each N210.
Each GPSDO outputs a \SI{10}{MHz} reference and a pulse-per-second (PPS)
signal, which provide frequency and time reference for all BSs~\cite{uhd2014synchronization}.
In addition, the UE is synchronized with the CBS through the air interface.
Therefore, we can achieve fine synchronization
between the UE, the associated CBS, and the serving TBS, which
serves as a base to provide a seamless service flow under the separation architecture.

\begin{table}[!t]
  \renewcommand{\arraystretch}{1.3}
  \caption{BSs and PCs in the HyCell testbed.}
  \label{tab:pc}
  \centering
  \begin{tabular}{lll}
    \hline
    BS& PC Model & Processor \\
    \hline
    CBS  & Dell Optiplex 990 & Intel Core i7-2600 \\
    TBS1 & Dell Optiplex 755 & Intel Core2 Duo  E6550 \\
    TBS2 & Dell Optiplex 9010 & Intel Core i7-3770 \\
    \hline
  \end{tabular}
\end{table}
\begin{figure*}[!t]
  \centering
  \subfloat{%
  \includegraphics[width=.32\textwidth]{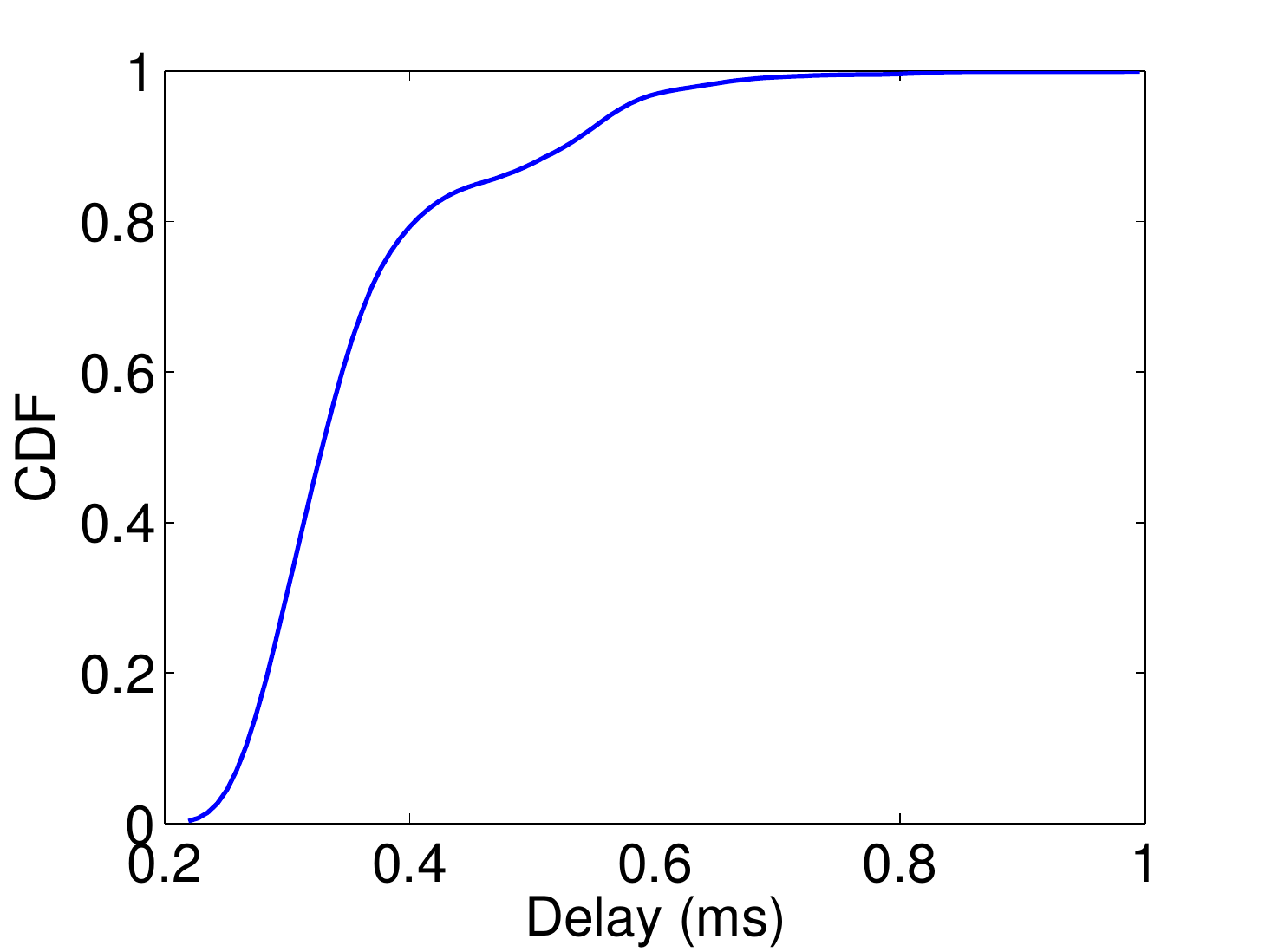}
  \label{fig:sig_cdf}}
  \hfil
  \subfloat{%
  \includegraphics[width=.32\textwidth]{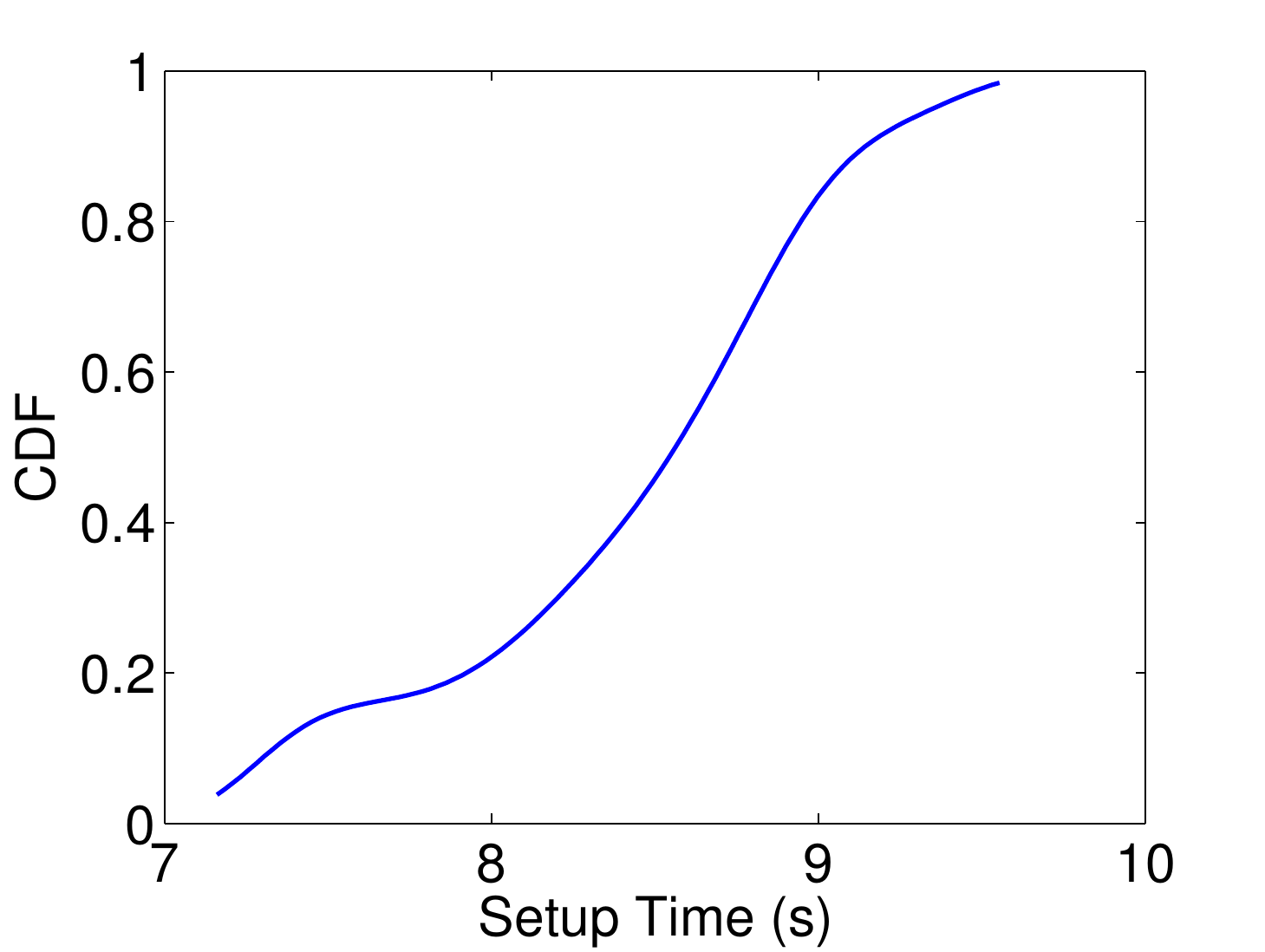}
  \label{fig:wakeup_cdf}}
  \hfil
  \subfloat{%
  \includegraphics[width=.32\textwidth]{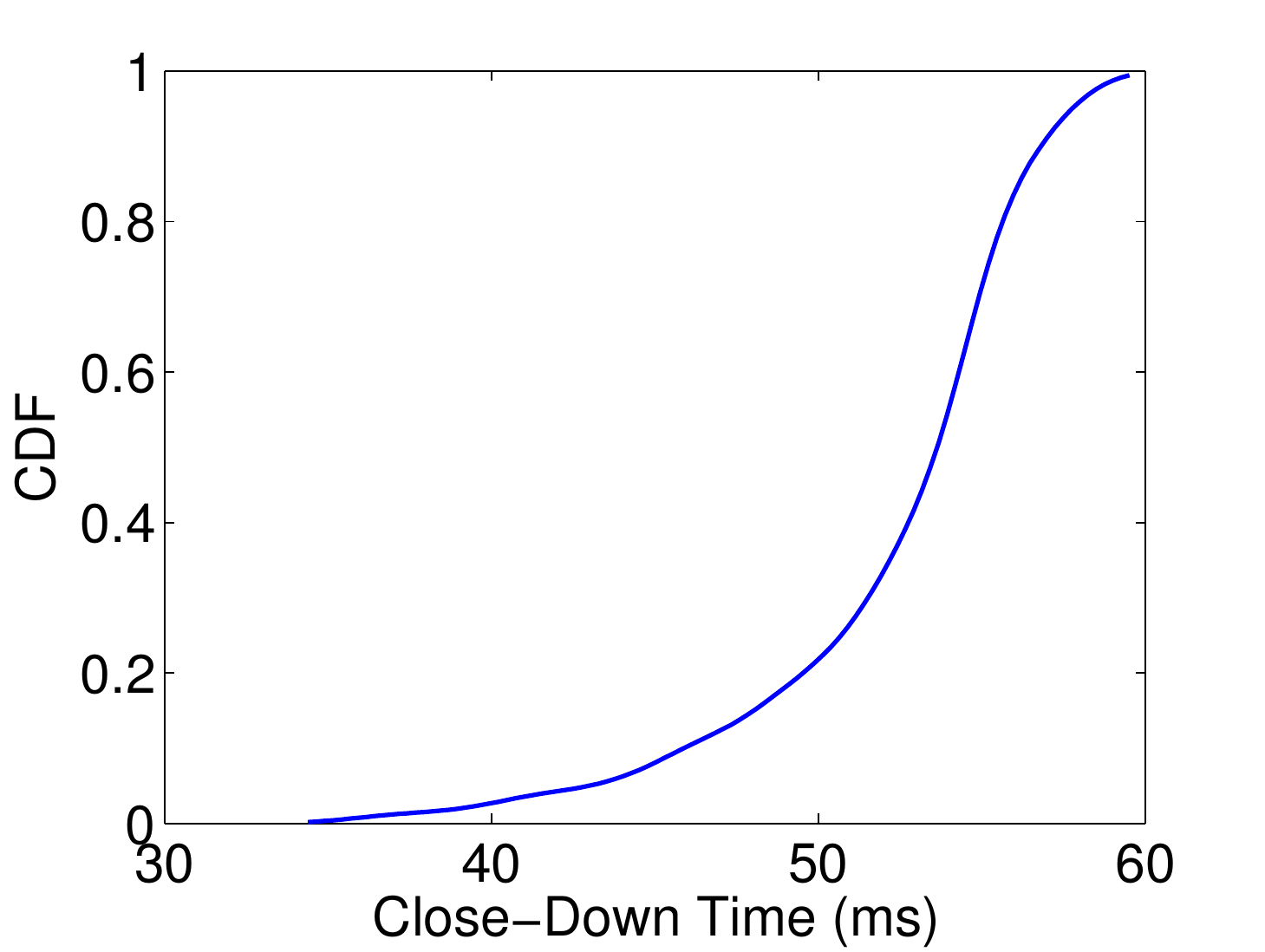}
  \label{fig:sleep_cdf}}
  \caption{CDF of (a) delay overhead, (b) setup time, and (c) close-down time.}
  \label{fig:cdf}
\end{figure*}

The BS applications are all running on Ubuntu GNU/Linux 12.04.1 LTS.
We developed our CBS and TBS applications based on a GPRS fork of OpenBTS
P2.8\footnote{\url{https://github.com/chemeris/openbts-p2.8}}
and Osmocom\footnote{\url{http://osmocom.org/}}
components including osmo-pcu, Osmo\-SGSN, and Open\-GGSN. Besides,
Asterisk\footnote{\url{http://www.asterisk.org/}} is used to connect phone calls.

\subsection{Feasibility of Separation}

We implement our proposed separation scheme of the GSM/GPRS air interface in the testbed to prove the
feasibility of separation.
In our tests, the CBS provides network coverage, voice call service, and short message service (SMS)
for UE, while UE can acquire GPRS data service from the TBSs.
Specifically, after the UE sends a channel
request to the CBS, the CBS analyzes the requested channel type. If the
channel type is PDTCH, the CBS dispatches a TBS to serve the UE.
The messages between the CBS and the TBS are transmitted through the wired
connection with UDP packets.
Afterwards, packet-switched data transmission is established between
the TBS and the UE.

In HyCell, the signaling interaction between the CBS and the TBS incurs an additional
delay when processing the channel request of high-rate data service. We term it as the
``delay overhead'' of separation. We measured the delay overhead in our
tests, and plot its distribution in Fig.~\ref{fig:sig_cdf}. The mean
value of the signaling overhead in our testbed is about \SI{0.36}{ms} and the
standard variance is about \SI{0.1}{ms}. The overhead is quite small and can be
negligible compared to the time that the UE waits before retransmitting the channel request.
Besides, in terms of user experience, there is no noticeable
service quality degradation in our tests.
Therefore, we validates the feasibility of
the separation scheme in our testbed.

\subsection{Effectiveness of Load Balancing}

We put the load-balancing dispatching scheme in Section~\ref{sec:disp} into practice and
evaluated its performance in the tests.
In the implementation we measure the quantity of
data transmitted on PDTCH and use it to calculate the traffic load of each TBS.
We do not include ARFCN usage in the packet in the testbed implementation
since each TBS only has one preconfigured radio carrier.

In our test scenario, initially a mobile user is served by
a single TBS. Then we activate another TBS and mobile phone.
At this time, the second TBS has no traffic load. Afterwards,
the CBS will assign the second TBS to
serve the channel requests of the new mobile user, which drives the load of the two TBSs to approach equal.
The results demonstrate that the scheme achieves load balancing of multiple TBSs.

\subsection{Base Station Sleeping}

We implement the protocol of TBS sleeping with a simple threshold-based
algorithm in our
testbed. In the implementation, sleeping and waking up of the TBSs are decided
by the CBS considering the total traffic load of the TBSs.
A sleeping and waking up module is created at each TBS, which receives the sleeping
or waking up command from the CBS, and turns off or turns on the TBS
by starting or stopping the TBS application accordingly.
As the on and off states of the USRPs cannot be controlled by the
baseband software, we need to power on or power off the USRPs manually.

We measured the setup and close-down time of BS sleeping, and show their
distributions
in Fig.~\ref{fig:wakeup_cdf} and Fig.~\ref{fig:sleep_cdf} respectively.
The average close-down time when turning off a TBS is \SI{52}{ms},
while the average setup time when waking up a TBS, during which the TBS
is initialized and synchronized, is \SI{8.4}{s}.
Note that the setup time is rather long, making the current BS sleeping and waking up
implementation unsuitable for services
that require low latency. Further work on reducing the setup time is thus
necessary.

\begin{table}[!t]
  \renewcommand{\arraystretch}{1.3}
  \caption{Power consumption comparison. (Unit: \si{W})}
  \label{tab:sav}
  \centering
  \begin{tabular}{c|c|c|c}
    \hline
    Component & No Sleep & Half Sleep &Full Sleep\\
    \hline
    PC & 346 & 310 &129 \\
    \hline
    USRP &41.4 & 13.8 & 13.8\\
    \hline
    Switch & 21 & 21 & 21\\
    \hline
    Total &408.4 &344.8 &163.8\\
    \hline
\end{tabular}
\end{table}

To show the energy saving performance of BS sleeping,
we currently take a component analysis approach due to the lack of power measurement equipment.
Based on the specifications,
we calculate the power consumption of PCs, USRPs and the switch respectively
and take the sum as the total power consumption of the testbed.
Table~\ref{tab:sav} compares the power consumption of different sleep modes in the
scenario when there are no active user requests.
Without BS sleeping (No Sleep in the table), all BSs are idle, and
the total power consumption is \SI{408.4}{W}.
When we only turn off the USRPs and software BSs on PCs (Half Sleep),
the total power consumption is \SI{344.8}{W}, that is, 16\% power consumption can be
saved.
If we further power off the PCs (Full Sleep), the total power can be reduced to \SI{163.8}{W},
which translates to 60\% energy savings.
It is thus observed that BS sleeping can bring significant energy saving gain,
of which the major part comes from the baseband power consumption.

\section{Conclusion}
\label{sec:con}

In this paper we present the design and evaluation of HyCell. The proposed
framework HyCell
exploits the decoupled air interface, centralized BS control, and software-defined BS
functions to enable GREEN BS operations in SDRANs. We propose
a separation scheme to realize the decoupled air interface, and
further design the protocols of BS dispatching and BS sleeping.
Our testbed implementation validates the feasibility of the separation scheme
for the GSM/GPRS standard.
Further the test results show that BS dispatching can effectively achieve load-balancing, and
BS sleeping can provide about 60\% energy saving gain.

\section*{Acknowledgment}

We would like to thank Yuyang~Wang and Guangchao~Wang for their help in
the evaluation. We would also like to thank Dr.~Xu~Zhang, Xueying Guo, Jingchu Liu,
and anonymous reviewers for their insightful comments.
This work is sponsored in part by the National Basic Research Program
of China (No. 2012CB316001), and the National Natural Science Foundation of
China (NSFC) under grant No. 61201191 and 61401250, the Creative Research Groups of
NSFC (No. 61321061), the Sino-Finnish Joint Research Program of NSFC
(No. 61461136004), and Intel Corporation.

\end{document}